\documentclass[aps,prd,twocolumn,superscriptaddress,amsfonts,amssymb,amsmath,eqsecnum,nofootinbib,floatfix]{revtex4}

\usepackage{graphicx,color,ulem}
\usepackage{subcaption}
\usepackage{algorithm}
\usepackage{algorithmicx}
\usepackage[noend]{algpseudocode}
\definecolor{darkblue}{rgb}{0,0,0.7}
\definecolor{darkred}{rgb}{0.7,0,0}
\usepackage[unicode, colorlinks, citecolor=darkblue, linkcolor=darkred, urlcolor=blue]{hyperref}

\newcommand{\beq}{\begin{equation}}
\newcommand{\eeq}{\end{equation}}
\newcommand{\bea}{\begin{eqnarray}} 
\newcommand{\eea}{\end{eqnarray}}

\algdef{SE}[DOWHILE]{Do}{doWhile}{\algorithmicdo}[1]{\algorithmicwhile\ #1}%

\allowdisplaybreaks

\definecolor{dgreen}{rgb}{.3,.7,.3}

\begin{document}
	\date{\today}
	
	\title{Gravitational Wave Driven Inspirals of Binaries Connected by Cosmic Strings}
	
	\author{Ahmed Sheta}
        % \email{asheta@g.harvard.edu}
        \affiliation{Department of Physics, Harvard University, Cambridge, MA 02138, USA}
	\affiliation{Department of Physics, Columbia University, New York, NY 10027, USA}
	\author{Yuri Levin}
	
	% \email{yl3470@columbia.edu}
	
	\affiliation{Department of Physics, Columbia University, New York, NY 10027, USA}
	\affiliation{ Center for Computational Astrophysics, Flatiron Institute, New York, NY 10010, USA}
	\affiliation{School of Physics and Astronomy, Monash Center for Astrophysics, Monash University, Clayton, VIC 3800, Australia}

\begin{abstract}
We consider gravitational waves from a pair of monopoles or black holes that are moving non-relativistically and are connected by a cosmic string. Shortly after the binary's formation, the connecting string straightens due the direct coupling of its motion to gravitational radiation. Afterwards, the motion of the binary can be well-approximated by a non-relativistic motion of its components that have an additional constant mutual attraction force due to the tension of the straight string that connects them. The orbit shrinks due to the gravitational radiation backreacting on the binary's components. We find that if the binary's semimajor axis $a\gg \sqrt{R_1 R_2/{\mu}}$, its eccentricity grows on the inspiral's timescale; here $R_1$ and $R_2$ are the gravitational radii of the binary components, and $\mu$ is the dimensionless tension of the string. When the eccentricity is high, it approaches unity super-exponentially. If the binary's components are monopole-antimonopole pair, this leads to the physical collision that would likely destroy the string and annihilate the monopoles when the semimajor axis is still many orders of magnitude greater than the string thickness. If the binary's components are black holes, then the eccentricity reaches its peak when $a\sim \sqrt{R_1 R_2/\mu}$, and then decays according to the standard Peter's formula. The black-hole spins initially become locked to the orbital motion, but then lag behind as the inspiral proceeds. We estimate the string-tension-induced dimensionless spins just prior to the merger and find them to be $\sim\mu^{3/8}\ll 1$. 

\end{abstract}

\pacs{}
\maketitle
\section{Introduction}
One of the widely considered possible outcomes of phase transitions in the early universe is the formation of networks of local Abelian cosmic strings \citep{1976JPhA....9.1387K, 2000csot.book.....V}. The laws governing the motions of such strings are extremely simple: they move according to the Nambu-Goto action and they reconnect on intersections with the probability of order $1$. These reconnections produce a multitude of oscillating loops that emit potentially detectable gravitational waves \citep{1981PhLB..107...47V, 2005PhRvD..71f3510D}. Despite the simplicity of the motion laws, understanding the evolution of the networks and calculating the distribution of the loop sizes presents a computational challenge, and only recently many groups started to consider these predictions as being reliable \cite{2020PhRvD.101j3018B}. Pulsar Timing Arrays measured an upper bound on the intensity of ambient gravitational waves passing through our Galaxy (the so-called Stochastic Gravitational Wave Background), which has placed a very interesting constraint on the dimensionless tension of the strings $\mu\lesssim 10^{-11}$, valid $\it if$ the string network indeed exists in our Universe \citep{2018PhLB..778..392B,2021PhRvD.103j3512B}. 

Interesting modifications of this scenario arise if multiple phase transitions take place. One of the common outcomes is the 
coexistence of strings and the monopoles that are attached to the string ends 
\citep{1980PhRvL..45....1L,1982NuPhB.196..240V, 2000csot.book.....V}. A common object that arises in such situation is a monopole-antimonopole pair connected by a string. The gravitational wave emission from such a binary was considered by Martin \& Vilenkin \cite{1997PhRvD..55.6054M}, but only in the limit of a straight string, with the monopoles on either a purely radial or purely circular orbit. The presence of such objects leads to substantial change in the expected spectrum of stochastic gravitational-wave background \citep{1996PhRvL..77.2879M,2021JCAP...12..006B,2022PhRvD.106g5030D}. Another interesting 
situation arises if primordial black holes are already present when the string network forms. Vilenkin et al. 
\cite{2018JCAP...11..008V} argued that in this case bound objects consisting of several black holes connected by strings can form, and the simplest such object is a pair of black holes connected by a string or a pair of strings (this is possible because black holes rapidly capture relativistically moving monopoles and antimonopoles).

The purpose of this work is to analyze the motion of such string-connected binaries with arbitrary angular momenta, and their evolution under the emission of gravitational waves. We consider a pair of objects (black holes or microscopic monopoles)  with gravitational 
radii $R_1$ and $R_2$, connected by a string of length $L$ with dimensionless tension $\mu\ll 1$. From this point onwards, we use geometrized units with $G=c=1$. The masses of the objects are then $m_{1,2}=R_{1,2}$. We assume that the string is much larger than the gravitational radii of the objects $L\gg R_1+R_2$, but that its mass is small compared to those of the objects, i.e. that $\mu L\ll R_1, R_2$. The case of a high-mass string is more complex and will be considered in a separate study. We do not assume {\it ab initio} that the string is straight. In Section \ref{sec:straightening} we will show, however, that gravitational radiation-reaction acting on the string will produce a non-relativistically moving binary connected by a nearly-straight string segment. The rest of the paper deals with the gravitational wave driven inspiral of such a binary, and is organized as follows. We review the inspiral in the gravity-dominated regime in Section \ref{subsec:gravity binary inspiral}; we study the inspiral in the string-dominated regime in Section \ref{subsec:string binary inspiral}; then we study the generic inspiral in Section \ref{subsec:general binary inspiral}. In Section \ref{sec:spins}, we study the evolution of the black-hole spins as the orbit inspirals and estimate the spins prior to mergers. We conclude with some discussion in Section \ref{sec:discussion}.

\section{Straightening of the sting} \label{sec:straightening}
We are interested in the regime where initially, the string tension is dominating the gravitational interaction between the objects, while the mass of the string is much smaller than that of either of the objects. Assume for definiteness that $R_2<R_1$. Therefore,
\begin{equation}
    \sqrt{R_1 R_2/\mu}\ll L \ll R_2/\mu.
\end{equation}
Given the likely smallness of $\mu$, there is a significant range of values of $L$ where these inequalities are satisfied.
The characteristic velocity of the smallest object is  $v_2\sim \sqrt{\mu L/R_2}$ and the characteristic orbital timescale is $P\sim L/v_2\sim \sqrt{L R_2/\mu}$.

The string, with its ends anchored on the heavy objects, oscillates and emits gravitational waves, losing its length in the process and straightening as a result. The timescale for this process can be estimated by assuming that the string is initially significantly non-straight and use the same 
scaling for the rate of length-loss as has been derived for free oscillating loops:
\begin{equation}
    \left(\frac{dL}{dt}\right)_{\rm GW}\sim -\Gamma \mu.
\end{equation}
The numerical factor $\Gamma$ is of order $50$ for free loops \cite{2000csot.book.....V}, but its exact value is unimportant and we can set it to $1$. The characteristic timescale 
on which the string straightens is given by
\begin{equation}
 t_{\rm straight}\sim L/\mu.
 \end{equation} 
A similar result can be obtained from computation of the decay time of a small perturbation of the straight string. The characteristic timescale for the string binary to shrink under gravitational radiation is given by
\begin{equation}
    t_{\rm GW}\sim R_2/\mu^2;
\end{equation}
see, e.g., Eq.~(30) of \cite{2018JCAP...11..008V} and Section \ref{subsec:string binary inspiral} of the current paper. From the equations above we see that
 \begin{equation}
   t_{\rm straight}/t_{\rm GW}\sim {\mu L/R_2}\ll 1.
 \end{equation}
Therefore, the string straightens on a short timescale compared to that of the binary inspiral. Hence, for the rest of the paper we shall assume that the binary members are connected by a straight string that creates an extra attractive force of magnitude $\mu$.

\section{Binary inspiral} \label{sec:binary inspiral}
The binary inspiral proceeds in two stages: first, when $a\gg \sqrt{R_1 R_2/\mu}$ and the attractive force between the binary components is dominated by the string tension, and second, when  $a\ll \sqrt{R_1 R_2/\mu}$ and the attractive force is entirely due to gravity. We shall refer to these stages as ``string-dominated inspiral" and ``gravity-dominated inspiral". Note that the second stage is not relevant for monopole-antimonopole pairs, since their gravitational attraction is extremely small and remains subdominant until the merger. In what follows we first review the gravity-dominated inspiral relevant for black holes; while this has been well-understood since Peters' work in $1964$, our treatment will set the stage for the techniques that we use to study the string-dominated inspiral, which we treat analytically. We then present numerical experiments that elucidate the transition from eccentricity-growing string dominated inspiral to eccentricity-damping gravity-dominated inspiral.
 
\subsection{Gravity-dominated inspiral.} \label{subsec:gravity binary inspiral}
In this sub-section, we review the classic results obtained in \cite{1964PhRv..136.1224P}. The methodology of our exposition will be useful in the analysis where the string tension is included.

The starting point of the analysis is the computation of the orbit-averaged rates of change of the orbital energy $E$ and angular momentum $\vec{J}$, given by
\begin{equation}
    \label{eq:eq1}
    \begin{aligned}
    & \left\langle \frac{dE}{dt} \right\rangle = - \frac{1}{5} \left\langle \frac{d^{3} I_{ij}}{dt^3}\frac{d^{3} I_{ij}}{dt^3} \right\rangle, \\
    & \left\langle \frac{dJ_{i}}{dt} \right\rangle = -\frac{2}{5} \epsilon_{ijk} \left\langle \frac{d^2 I_{jm}}{dt^2} \frac{d^3 I_{km}}{dt^3} \right\rangle.
    \end{aligned}    
\end{equation}
Here $\langle \rangle$ stand for averaging over an orbital period, and $I_{ij}$ is the traceless part of the quadrupole moment. Repeated indices are implicitly summed over in our notationa. These equations specify secular evolution of the orbit due to gravitational wave emissions. When it's clear from the context, we will omit $\langle \rangle$ to avoid notational clutter, such that ${dO}/{dt}$ implicitly refers to $\left\langle {dO}/{dt} \right\rangle$ for observable $O$. For a Keplerian orbit, one gets 
\begin{equation}
    \begin{aligned}
        & \frac{dE}{dt} = -\frac{32}{5} \cdot \frac{(R_1+R_2)(R_1R_2)^2}{a^5} \cdot f_E^G(e) \\
        & \frac{dJ}{dt} = -\frac{32}{5} \cdot \frac{(R_1+R_2)^{1/2}(R_1R_2)^2}{a^{7/2}} 
        \cdot f_J^G(e) \label{evolution1}
    \end{aligned}
\end{equation}
where $f_E^G(e)$ and $f_J^G(e)$ are given by
\begin{align*}
    & f_E^G(e) = \frac{1+\frac{73}{24} e^2 + \frac{37}{96} e^4}{ (1-e^2)^{7/2}} \\
    & f_J^G(e) = \frac{1+\frac{7}{8}e^2}{(1-e^2)^2}
\end{align*}

\begin{figure}[H]%
    \centering
    \includegraphics[width=8.5cm]{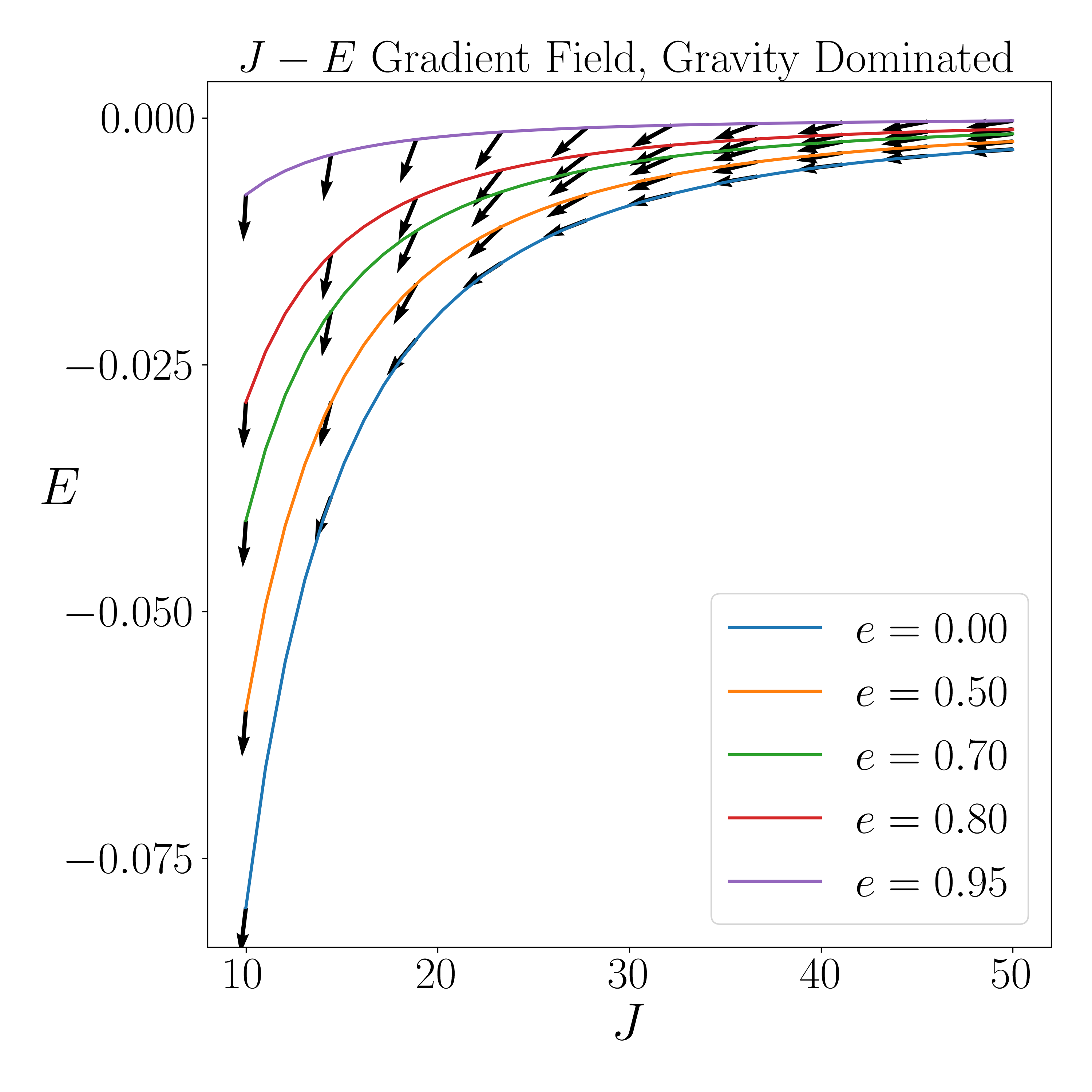}
    \caption{The flow field, $\left(\frac{dJ}{dt}, \frac{dE}{dt} \right)$, due to gravitational wave emission. The lengths of the vectors are non-uniformly scaled down for readability, but the angles are preserved. In particular, the leftmost vectors (with the lowest $J$) are scaled down by a factor of $\sim 10^6$, while the rightmost vectors with the highest $J$ are approximately drawn to scale. The curves represent constant eccentricity contours in the $J-E$ phase space. We verified our numerical results, obtained using the method outlined in Section \ref{subsec:string binary inspiral}, against Peter's formulae of Eq (\ref{evolution1}), and found agreement within numerical errors of order $10^{-4}$. Here $E$ and $J$ are measured in units of $R$ and $R^2$ respectively, where $R=R_1R_2/(R_1+R_2)$ is the reduced mass of the binary.}
    \label{fig:kep EL decay}%
\end{figure}

Using $E = -{R_1 R_2}/{(2a)}$ and $J = (R_1R_2)^2(R_1+R_2)^{-1/2}~\sqrt{a~(1-e^2)}$, one obtains the decay rate for the semimajor axis $a$ and the eccentricity $e$, as follows
\begin{equation}
    \label{eq:kep ae rates}
    \begin{aligned}
    & \frac{da}{dt} = - \frac{64}{5} \cdot \frac{(R_1+R_2)R_1 R_2}{a^3} \cdot \frac{1+\frac{73}{24} e^2 + \frac{37}{96} e^4}{ (1-e^2)^{7/2}}, \\
    & \frac{de}{dt} = -\frac{304}{15} \cdot \frac{(R_1+R_2) R_1 R_2}{a^4} \cdot \frac{e+\frac{121}{304}e^3}{(1-e^2)^{5/2}}.
    \end{aligned}
\end{equation}

The resulting evolution features a polynomial decay of the semi major axis, ${a^4}/{a_0^4} \sim 1 -{t}/{\tau}$, with the characteristic timescale $\tau \sim {a_0^4}/[(R_1+R_2)R_2 R_2]$, where $a_0$ is the initial semimajor axis. Similarly, the eccentricity also decays polynomially in time, with the same characteristic timescale of decay as the semi major axis. The result is that an arbitrarily eccentric orbit always approaches a circular orbit asymptotically as it decays. In fact, the secular evolution of the orbit can be solved exactly to obtain
\begin{equation}
    \label{eq:a(e)}
    a(e) = c_0 \frac{e^{12/19}}{1-e^2} \left( 1+\frac{121}{304}e^2 \right)^{870/2299},
\end{equation}
where $c_0$ is a constant determined from $a(e_0)=a_0$, $e_0$ being the initial eccentricity.
The well-known corollary of this equation is that $e \to 0$ as $a \to 0$, i.e.~the orbit circularizes as the inspiral proceeds. This circularization has a huge impact on gravitational-wave astronomy, and explains why the vast majority of LIGO mergers are circular.

It is instructive to visualize the eccentricity evolution in a $J-E$ plane, as shown in Fig.~(\ref{fig:kep EL decay}). The evolution equations (\ref{evolution1}) define a flow field for motion in the plane, shown by arrows in the figure. Solid lines represent curves of constant eccentricity, and the flow takes eccentric orbits towards the circular $e=0$ orbit. The decrease of eccentricity takes place for motion in a Keplerian potential, but it is not universal for all attractive potentials. As we show in the next subsection, if the attractive force is distance-independent, the eccentricity increases during the inspiral.

\subsection{String-dominated inspiral} \label{subsec:string binary inspiral}
In this section we consider the case where the binary is bound together by a straight cosmic string, and the influence of gravitational attraction can be neglected. 

\paragraph{Orbital Mechanics}
 For an orbit with angular momentum $J$, the effective radial potential is given by
\begin{equation}
    U_J(r) = \mu r + \frac{J^2}{2Rr^2};
    \label{potential}
\end{equation}
here $r$ is the distance between the centers of mass of the binary components, and $R=R_1R_2/(R_1+R_2)$ is the reduced mass of the binary in geometrized units.
The rosette-like orbit is no longer closed, but the radial motion is periodic. Let $r_a$ and $r_p$ be the maximum and minimum values of $r$ during the motion. Lets define, in analogy with Keplerian orbits, the semimajor axis and the eccentricity as follows:
\begin{eqnarray}
    a&\equiv&(r_a+r_p)/2\nonumber\\
    e&\equiv& \frac{r_a-r_p}{r_a+r_p}.
\end{eqnarray}
By using $U(r_p)=U(r_a)=E$, it is straightforward to obtain the following expressions for the energy and the angular momentum of the binary:
\begin{equation}
    \label{eq:k=0, El}
    \begin{aligned}
        & E=\frac{3}{2} \cdot \mu a \cdot (1+e^2/3) \\
        & J=\sqrt{\mu R a^3} \cdot (1-e^2)
    \end{aligned}
\end{equation}
The constant eccentricity curves in the $J-E$ phase space can be specified as follows:
\begin{equation}
    \label{eq: k=0 E(l)}
    E_e(J) = \frac{\mu^{2/3}}{2R^{1/3}} \cdot  \frac{3+e^2}{(1-e^2)^{2/3}} \cdot J^{2/3}.
\end{equation}
This is quite different from the relationship for Keplerian orbits, $E_e(J) = -(R_1+R_2)^2 R^3~(1-e^2)/({2J^2})$. Note that for Keplerian orbits, the decay rate of $e$ depends on the evolution of $E  J^2$, whereas in the pure string-tension-dominated orbits the evolution of $e$ depends on that of $E J^{-2/3}$.

The radial velocity is given by
\begin{eqnarray}
v_r&\equiv& \frac{dr}{dt}=\pm [2(E-U_J)/R]^{1/2}\nonumber\\
 &=&\pm \sqrt{2\mu\over R}{\left[(r_a-r)(r-r_p)(r+r_3)\right]^{1/2}\over r},
 \label{vrad}
 \end{eqnarray}
where 
\begin{eqnarray}
    r_a&=&a(1+e),\nonumber\\
    r_p&=&a(1-e),\nonumber\\
    r_3&\equiv& {1\over 2}a(1-e^2).
\end{eqnarray}
The period $P$ of the radial motion is given by
\begin{eqnarray}
P(a,e)&=&2\int_{r_p}^{r_a} |v_r|^{-1} dr\nonumber\\
 &=&2\sqrt{a R\over\mu}{1\over\sqrt{3+2e-e^2}}\times\nonumber\\
 & &\left[(e^2-1)K\left({4e\over 3+2e-e^2}\right)+\right.\label{period1}\\
 & &\left.(3+2e-e^2)E\left({4e\over 3+2e-e^2}\right)\right]\nonumber.
\end{eqnarray}
Here $K$ and $E$ are the complete elliptic integrals of the first and second kind, respectively (``EllipticK" and ``EllipticE" in Mathematica\footnote{There is a slight discord in the literature on the precise definition of these functions. We follow here the definitions used in Matlab and Mathematica, namely: 
\newline
$K(m)=\int_0^{\pi/2}(1-m \sin^2\theta)^{-1/2} d\theta$,\newline  $E(m)=\int_0^{\pi/2}(1-m \sin^2\theta)^{1/2} d\theta$,\newline  $\Pi(n,m)=\int_0^{\pi/2}(1-n\sin^2\theta)^{-1}(1-m\sin^2\theta)^{-1/2} d\theta$}). For the extreme values of eccentricity we have
\begin{eqnarray}
P(a,0)&=&2\pi \sqrt{aR\over 3\mu},\nonumber\\
P(a,1)&=&4\sqrt{aR\over \mu}.
\end{eqnarray}
The former corresponds to the period of small-amplitude oscillation about the circular orbit of distance $a$ between the binary members. The latter corresponds to the time interval for the purely radial motion between two sequential full stops, with $2a$ being the maximal distance between the binary members. 

Some relations for orbit-averaged quantities can be derived that will prove useful in the computation of gravitational-wave emission:
\begin{eqnarray}
 \left\langle r\right\rangle&=& a(1+e^2/3),\label{rel1}\\
 \left\langle {1\over r}\right\rangle&=&{2\over (1+e)} K\left({4e\over 3+2e-e^2}\right)\times\nonumber\\
  & &\left[(e-1)K\left({4e\over 3+2e-e^2}\right)+\right.\label{rel2}\\
 & &\left.(3-e)E\left({4e\over 3+2e-e^2}\right)\right]^
 {-1}{1\over a},\nonumber\\
 \left\langle {1\over r^2}\right\rangle&=&{2\over (1+e)^2} \Pi\left({2e\over e+1},{4e\over 3+2e-e^2}\right)\times\nonumber\\
  & &\left[(e-1)K\left({4e\over 3+2e-e^2}\right)+\right.
  \label{rel3}\\
 & &\left.(3-e)E\left({4e\over 3+2e-e^2}\right)\right]^
 {-1}{1\over a^2},\nonumber\\
 \left\langle {1\over r^3}\right\rangle&=&{1\over a^3 (1-e^2)^2}.\label{rel4}
\end{eqnarray}
Here $\Pi$ is the complete elliptic integral of the third kind (``EllipticPi'' in Mathematica). The derivation of these relations is sketched in Appendix \ref{sec:app averages}.

\paragraph{Orbital Decay.}
First, lets consider the decay of a circular orbit given by
\begin{align*}
    \vec{r} = a \left[\mathrm{cos}(\omega t), \mathrm{sin}(\omega t)\right]
\end{align*}
where $\omega=\sqrt{\mu/(Ra)}$. The non-zero terms of the traceless quadrupole tensor are given by
\begin{equation}
    \begin{aligned}
    & I_{xx} = R a^2 \left[\mathrm{cos}^2(\omega t) - \frac{1}{3} \right]\\
	& I_{yy} = R a^2 \left[\mathrm{sin}^2(\omega t) - \frac{1}{3} \right]\\
	& I_{zz} = -\frac{1}{3} R a^2 \\
	& I_{xy} = R a^2 \mathrm{cos}(\omega t) \mathrm{sin}(\omega t)
    \end{aligned}
    \label{Icirc}
\end{equation}
We use Eq (\ref{eq:eq1}) to calculate the rate of change in $E$ and $J$:
\begin{equation}
    \begin{aligned}
    & \frac{dE}{dt} = -\frac{32}{5} \frac{\mu^3}{R} a \\
    & \frac{dJ}{dt} = -\frac{32}{5} \frac{\mu^{5/2}}{R^{1/2}} a^{3/2}
    \end{aligned}
\end{equation}
For a circular orbit, $E=(3/2)\mu a$. We therefore obtain  ${da}/{dt} = -({64}/{15}) {\mu^2} a/R$, 
and
\begin{equation}
     a(t) = a_0 \exp\left({-\frac{64 }{15} \frac{\mu^2}{R} t}\right).
\label{decay1}
\end{equation}

Another limiting case is periodic radial oscillation considered by \cite{1997PhRvD..55.6054M}. Newton's equations give ${x(t) = r_0 -({\mu}/{2R}) t^2}$, valid for half a period of a particle that starts at rest along the $x-$axis at $r_0 =  2a$. The non-zero components of the traceless quadrupole tensor are
\begin{equation}
    \begin{aligned}
        &I_{xx} = \frac{2}{3} R x^2, \\
        &I_{yy} = I_{zz} = - \frac{1}{3}R x^2,
    \end{aligned}
\end{equation}
which we substitute in Eq (\ref{eq:eq1}) to obtain 
\begin{equation}
\begin{aligned}
    & \frac{dE}{dt} = -\frac{32}{5} \frac{\mu^3}{R} a, \\
    & \frac{dJ}{dt} = 0.
\end{aligned}
\end{equation}
Using $E = 2 \mu a$, we obtain
\begin{equation}
a(t) = a_0 \exp\left({-\frac{16}{5} \frac{\mu^2}{R} t}\right).
\end{equation}

For arbitrary $e$, we expect that
\begin{equation}
    \begin{aligned}
    & \frac{dE}{dt} (e) = \frac{dE}{dt} (0) \cdot f_E^\mu(e) = -\frac{32}{5} \frac{\mu^3}{R} a \cdot f_E^\mu(e) \\
    & \frac{dJ}{dt} (e) = \frac{dJ}{dt} (0) \cdot f_J^\mu(e) = -\frac{32}{5} \frac{\mu^{5/2}}{R^{1/2}} a^{3/2} \cdot f_J^\mu(e)
    \end{aligned}
    \label{eccentricityfactors}
\end{equation}
for the form factors $f_E^\mu(e)$ that varies smoothly between $f_E^\mu(0) = 1$ and $f_E^\mu(1) = 1$, and $f_J^\mu(e)$ that varies smoothly between $f_J^\mu(0) = 1$ and $f_J^\mu(1) = 0$. The functions $f_E^\mu(e)$ and $f_J^\mu(e)$ are computed analytically in Appendix \ref{sec:app fE_fJ}; one obtains 
\begin{eqnarray}
f_E^\mu(e)&=& {(1-e)^2\over 2} \Pi\left({2e\over e+1},{4e\over 3+2e-e^2}\right)\times\nonumber\\
 & &\left[(e-1)K\left({4e\over 3+2e-e^2}\right)+\right.\label{eq:fE}\\
 & &\left.(3-e)E\left({4e\over 3+2e-e^2}\right)\right]^{-1}+{3+e^2\over 4},\nonumber\\
\label{eq:fJ}
 f_J^{\mu}(e)&=&{1-e^2\over 4}\left\{1+2(3+e^2)K\left({4e\over 3+2e-e^2}\right)\times\right.\nonumber\\
  & &\left[(e^2-1)K\left({4e\over 3+2e-e^2}\right)+\right.\\
  & &\left.\left. (3+2e-e^2)E\left({4e\over 3+2e-e^2}\right)\right]^{-1}\right\}\nonumber
\end{eqnarray}
These results have been checked by numerically integrating the equations of motion using the effective potential in Eq.~(\ref{potential}) and evaluating the derivatives of the quadrupole moments in Eq.~(\ref{eq:eq1}). The agreement between the analytical and numerical calculations is very good; the results are plotted in Fig.~(\ref{fig:fE,fl}).

\begin{figure}[H]%
    \centering
    \subfloat{{\includegraphics[width=8.5cm]{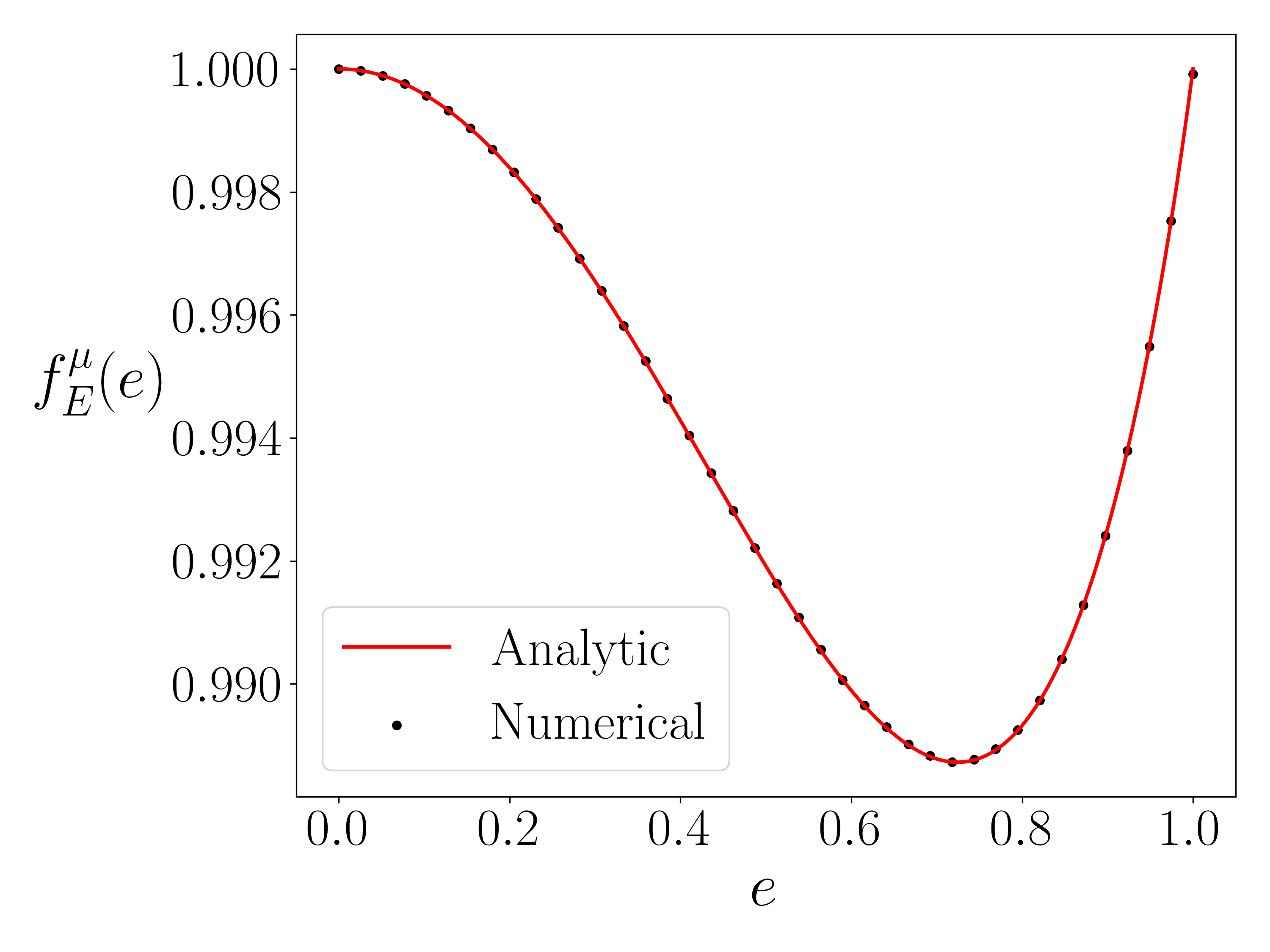} }}%
    \qquad
    \subfloat{{\includegraphics[width=8.5cm]{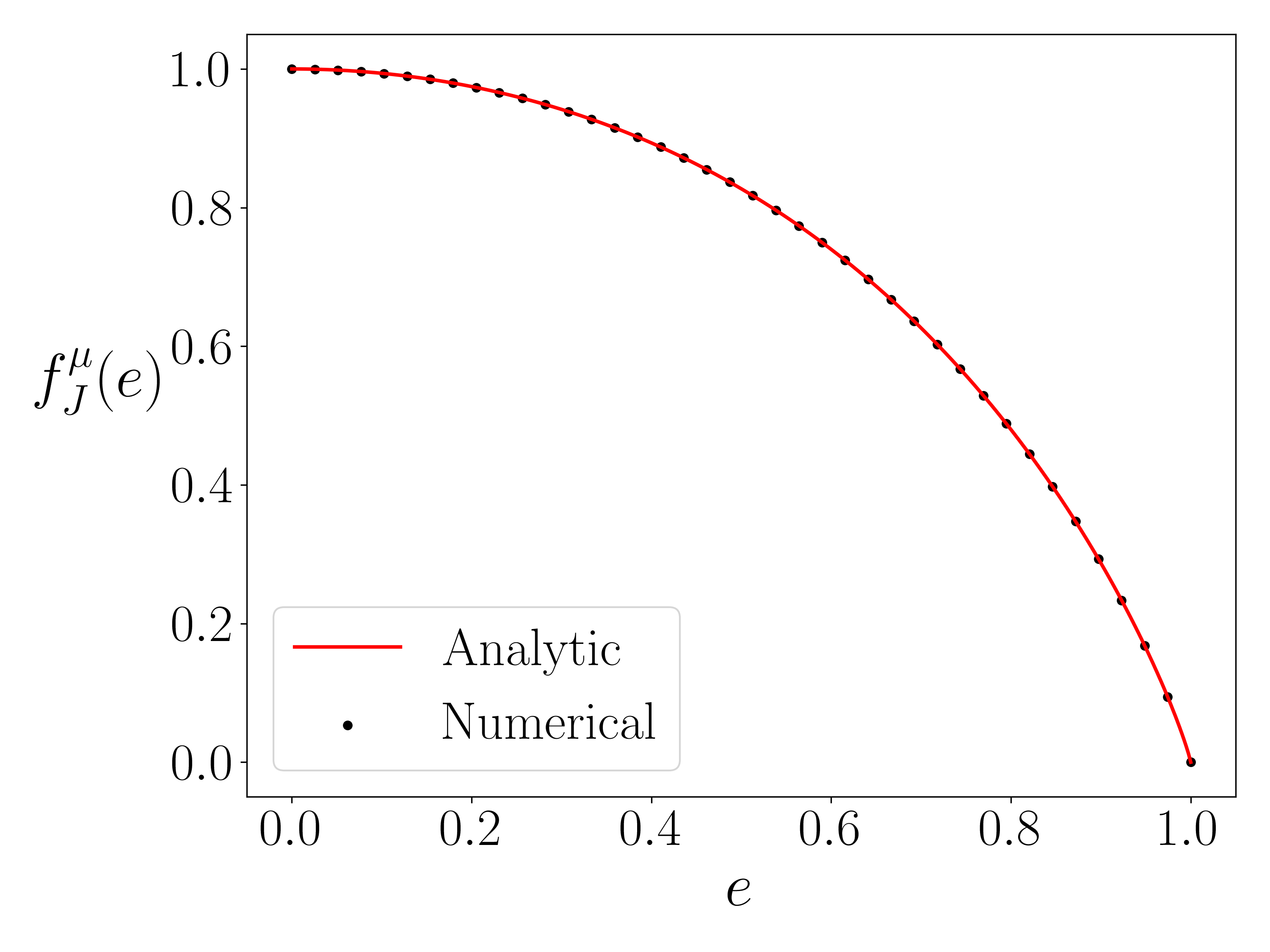} }}%
    \caption{Eccentricity-dependent form factors in the $J,E$ evolution equations (\ref{eccentricityfactors}). The numerical computation is verified against the analytic predictions, and was found to be accurate within $10^{-5}$.}
    \label{fig:fE,fl}%
\end{figure}

Using Eq (\ref{eq: k=0 E(l)}), we have
\begin{equation}
    \frac{3+e^2}{(1-e^2)^{2/3}} = \frac{2E R^{1/3}}{(\mu J)^{2/3}}
\end{equation}
which we use to obtain the evolution rate of the eccentricity:
\begin{eqnarray}
    \label{eq: dedt k=0}
    \frac{de^2}{dt}&=&\frac{64}{15} \cdot \frac{\mu^2}{R} \cdot \frac{(1+e^2/3) \cdot f_J^\mu(e) - (1-e^2) \cdot f_E^\mu(e)}{1-e^2/9}\nonumber\\
    &\equiv& \frac{64}{15} \cdot \frac{\mu^2}{R} \cdot g_e(e)
\end{eqnarray}
where $g_e(e)$ is plotted in Fig.~(\ref{fig:ge,ga}). We note that ${de^2}/{dt} > 0$ for all $0<e<1$. Therefore an eccentric binary increases its eccentricity during the string-dominated stage of the inspiral. 

For small eccentricity $e\ll1$, we can expand the expression above to second order in $e$. (We used the Series$[{\rm ``expression"
},\{e,0,2\}]$ command in Mathematica. It was expedient as an intermediate step to first obtain the bivariate expansion of $\Pi(x,y)$ to second order in $x,y$, and then enter it in place of $\Pi$ in the expression for $f_E^{\mu}(e)$.) To the lowest order in $e$ we obtain
a simple answer $g_e(e)= (3/4)e^2$ which we cross-check numerically. Thus a small eccentricity grows exponentially with time
\begin{equation}
e(t)=e_0\exp\left({8\over 5}{\mu^2\over R}t\right).
\label{eccgrowth}
\end{equation}
and as a power-law with the inverse semimajor axis:
\begin{equation}
e(t)=e_0\left[{a(t)\over a_0}\right]^{-3/8}.
\end{equation}
The non-negligible increase in eccentricity takes place on roughly the same time scale as that of the orbital decay, and an orbit that starts with very minimal eccentricity $e_0 \ll 1$ will evolve to $O(1)$ eccentricity as it inspirals.

Likewise, for a nearly radial orbit $e = 1-\delta$ with $\delta \ll 1$, we expand Eq (\ref{eq: dedt k=0}) to leading-order in $\delta$, and we obtain $g_e(1-\delta) =(3 \ln 2 - 3/2) \delta -(3/4) \delta \ln \delta$, which we also cross-check numerically. The evolution equation for 
$\ln\delta$ is, approximately,
\begin{equation}
    {d\ln\delta\over dt}={8\mu^2\over 5R}[\ln\delta-4(\ln 2-0.5)].
\end{equation}

Thus, a nearly radial orbit approaches periodic radial behavior super-exponentially with time. To reach even a tiny $\delta_f \sim \exp(-N)$, where $N\gg 1$, takes
only time
\begin{equation}
t_f\simeq{5R\over 8\mu^2}\ln N.
\label{superexponential}
\end{equation}
The above equation, together with Eq.~(\ref{eccgrowth}) implies that in the string-dominated inspiral the objects will physically collide only after time $t_f=\hbox{few}\times 5R/(8\mu^2)$. For example, if the initial size of the binary is $\sim 1$pc and the sizes of the objects or the physical string width are $1$fm, the required $\delta$ for physical collision is ${\delta_f\sim 10^{-30}\sim \exp(-69)}$, and so $t_f\simeq 4\times 5R/(8\mu^2)$. For sufficiently heavy black holes, however, the Peters' inspiral regime will take over before the objects will physically collide and the orbit will circularize before the merger.

To determine the evolution of $a$ for general eccentricity, we use Eq (\ref{eq:k=0, El}) and find 
\begin{equation}
    E = \frac{1}{2} \mu a \left( 4 - \frac{J}{\sqrt{\mu R a^3}} \right).
\end{equation}
Differentiating this with respect to time and using Eqs.~(\ref{eccentricityfactors}), we obtain
\begin{eqnarray}
    \label{eq: dadt k=0}
    \frac{da}{dt}&=& -\frac{64}{15} \frac{\mu^2}{R} a \cdot \frac{2f_E(e) + f_J(e) }{3(1-e^2/9)}\nonumber\\
    &\equiv& -\frac{64}{15} \frac{\mu^2}{R} a g_a(e).
    \label{aevolution}
\end{eqnarray}
The function $g_a(e)$ is plotted in Fig.~(\ref{fig:ge,ga}).
\begin{figure}[H]%
    \centering
    \subfloat{{\includegraphics[width=8.5cm]{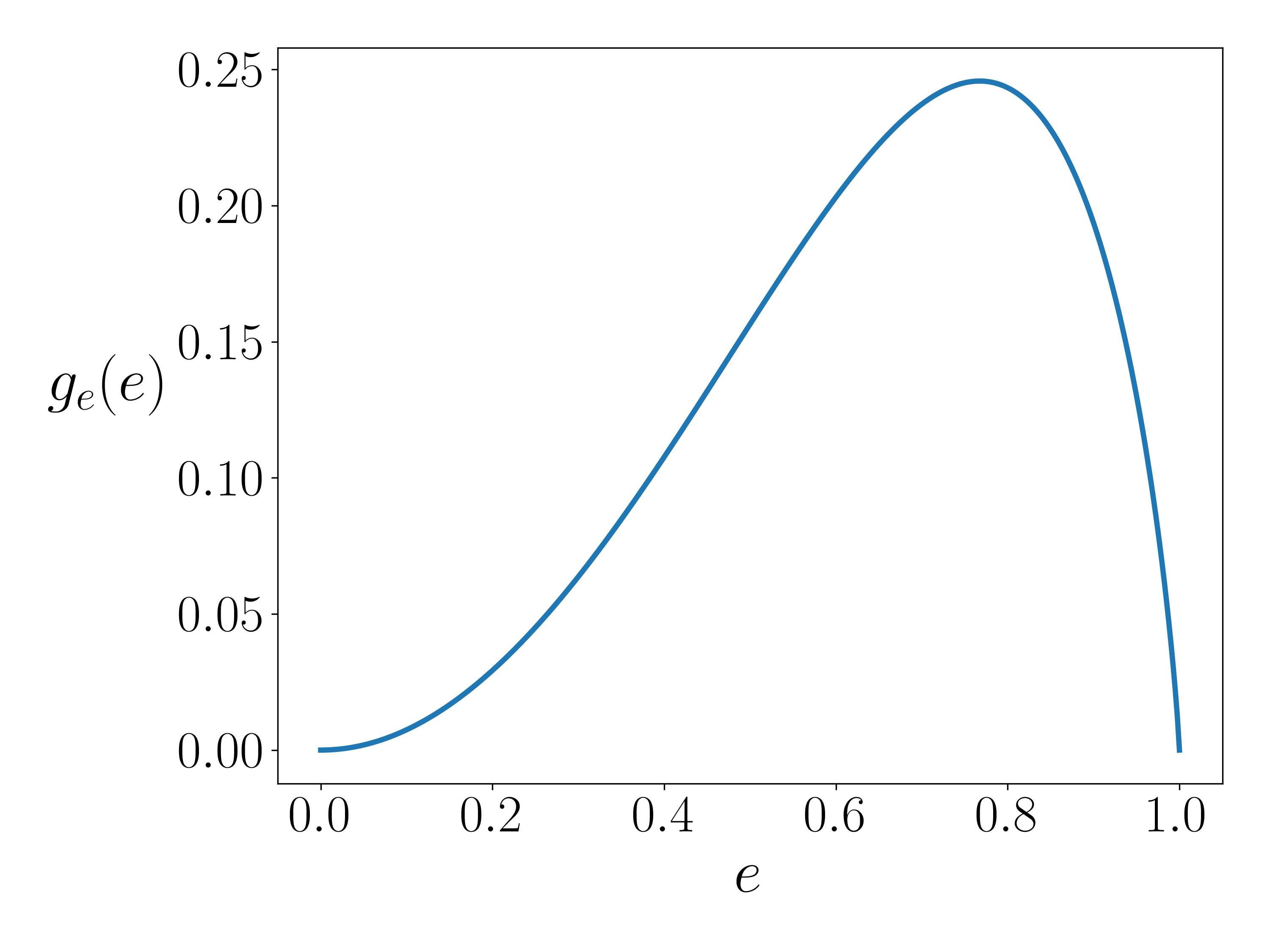} }}%
    \qquad
    \subfloat{{\includegraphics[width=8.5cm]{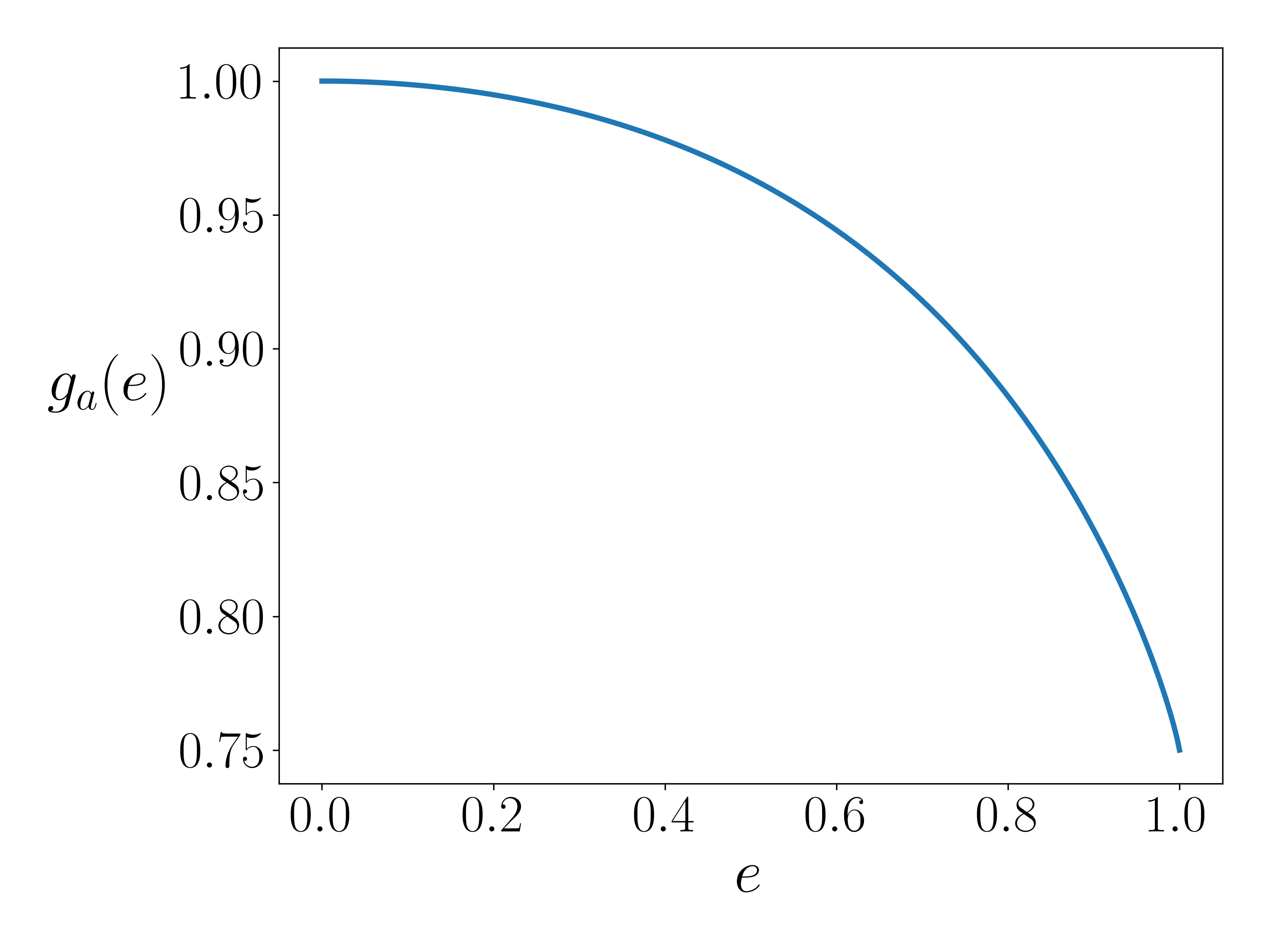} }}%
    \caption{Eccentricity factors in the radiation rates of $e,a$.}
    \label{fig:ge,ga}%
\end{figure}

\begin{figure}[H]%
    \centering
    \includegraphics[width=8.5cm]{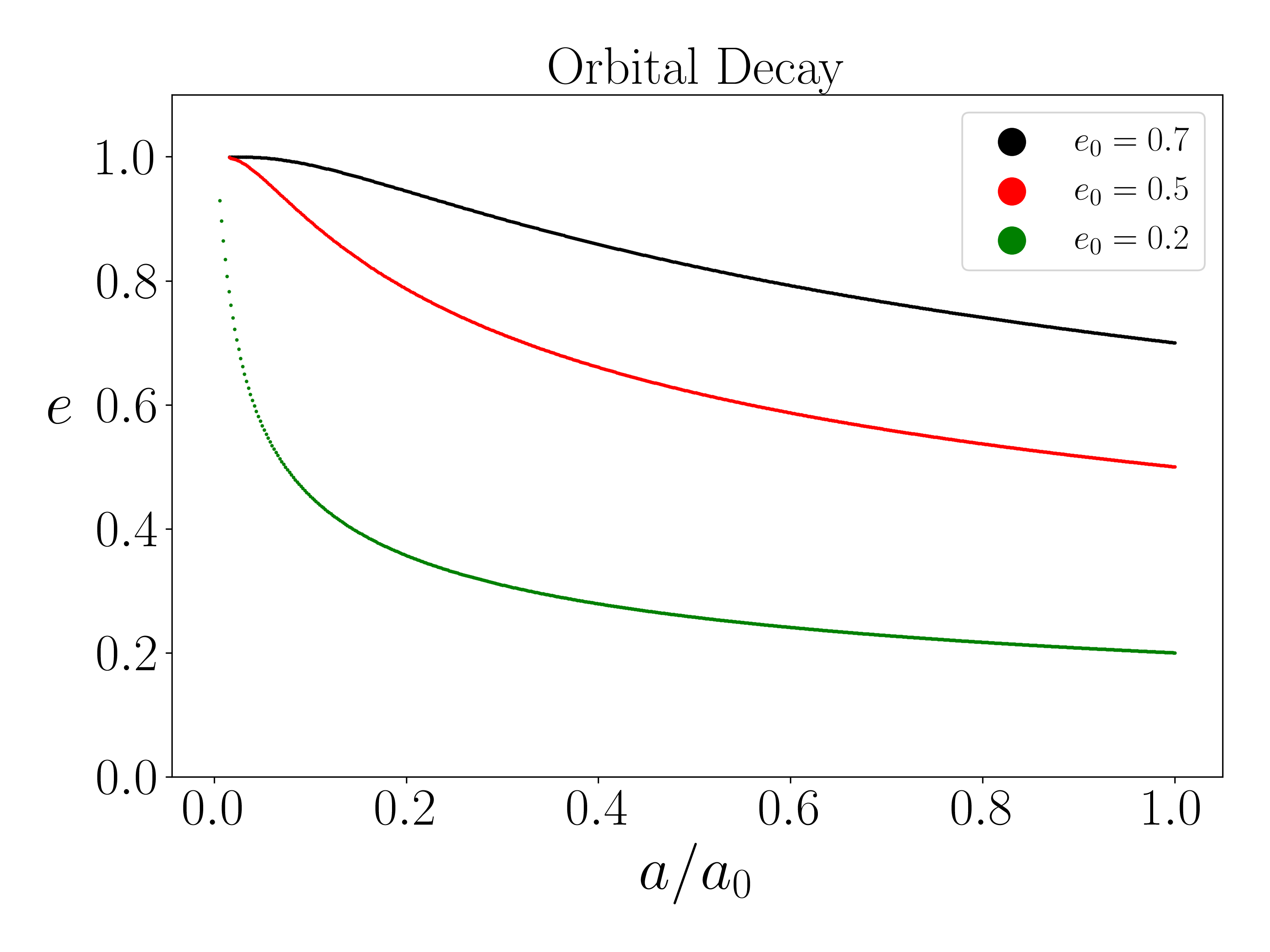}
    \caption{Numerical calculation of the decay under pure string tension, until $a(t)/a_0 < 1/150$.}
    \label{fig:k0 decay}%
\end{figure}

Dividing Eq.~(\ref{aevolution}) by Eq.~(\ref{eq: dedt k=0}) and solving the resulting differential equation, we obtain the general relationship between $a$ and $e$ for the inspiraling orbit:
\begin{equation}
    \label{eq: k=0 a(e)}
    a(e) = a_0 \exp{\left( - 2 \int_{e_0}^e \frac{e' \cdot g_a(e')}{g_e(e')} de' \right)}
\end{equation}
Plugging in Eqs.~\eqref{eq:fE} and~\eqref{eq:fJ} into $g_e(e), g_a(e)$, defined in Eqs.~\eqref{eq: dedt k=0} and~\eqref{eq: dadt k=0} respectively, we get an analytic expression for the secular evolution, but it's hard to integrate in closed form. Instead, several numerical examples of this relationship are plotted in Fig.~(\ref{fig:k0 decay}).

\begin{figure}[H]%
    \centering
    \includegraphics[width=8.5cm]{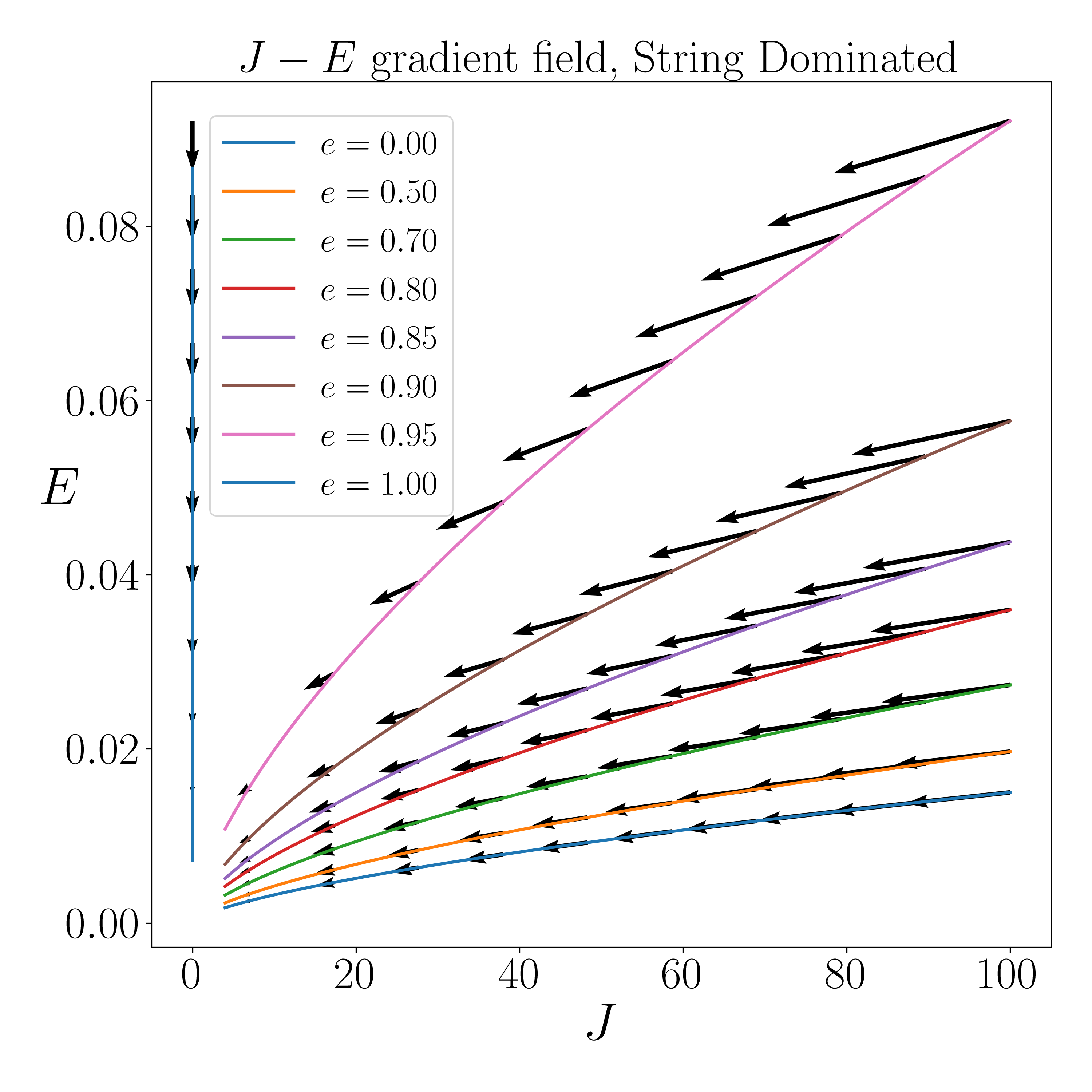}
    \caption{Gradient field, $\left(\frac{dJ}{dt}, \frac{dE}{dt} \right)$, due to gravitational wave radiation, with $\mu = 10^{-5}$. The lengths of the vectors are uniformly scaled up by a factor of $5 \times 10^{9}$, but their angles are preserved. Here $E$ and $J$ are measured in units of $R$ and $R^2$ respectively, where ${R=R_1R_2/(R_1+R_2)}$ is the reduced mass of the binary.}
    
    \label{fig:k=0 EL decay}%
\end{figure}

It is instructive to consider the gradient field in the ${J-E}$ plane, plotted for some sample orbits in Fig.~(\ref{fig:k=0 EL decay}). We see that the vector field monotonically flows towards higher eccentrcitiy, except for the two special cases $e=0$ and $e=1$ for which the flow is along the constant eccentricity curves. This agrees with the preceding discussion.

In contrast with the Keplerian case, gravitational wave from the string-driven inspiral does not produce much of a chirp.  The characteristic frequency $\omega_0 \propto \sqrt{\frac{1}{a}}$ increases exponentially with time, while the power of the emitted gravitational waves decreases. For very eccentric orbits we do not expect any sharp features since the acceleration remains very regular, in a distinction from conventional eccentric inspirals where the pericenter passages are accompanied by strong gravitational-wave bursts.

\subsection{General case} \label{subsec:general binary inspiral}
In this sub-section, we consider the general case where both gravitational attraction and cosmic string tension are included in the potential. First, we briefly discuss the orbital mechanics without radiation. Then, we consider the inspiral of a circular orbit, which can be exactly solved analytically but which is unstable to eccentricity growth in the string-dominated part of the inspiral. Finally, we employ numerical methods to obtain the inspiral of a general eccentric orbit.

\begin{figure}[H]%
    \centering
    \includegraphics[width=8.5cm]{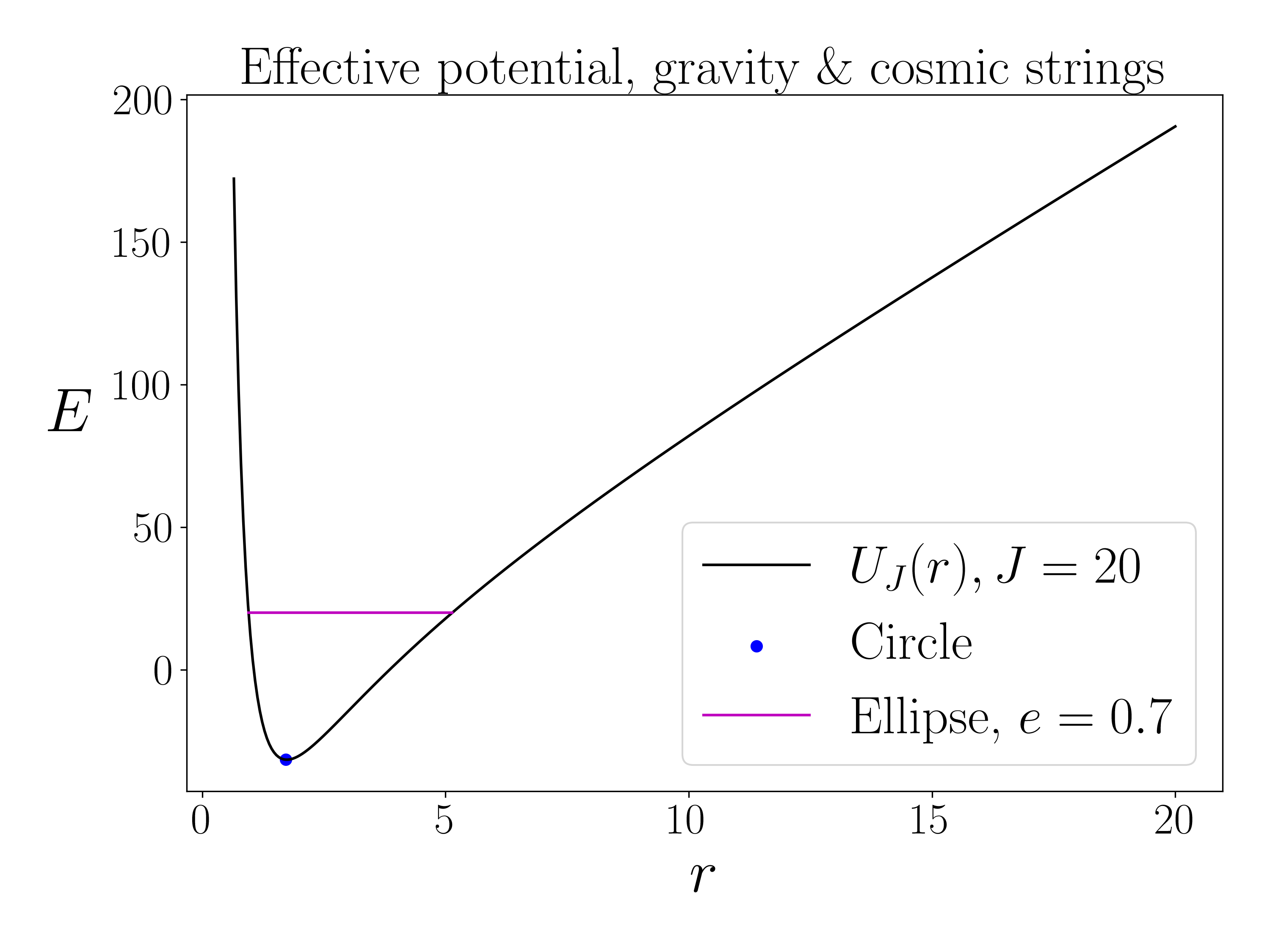}
    \caption{Here $r$ and $E$ are measured in units of $R=R_1R_2/(R_1+R_2)$, the reduced mass of the binary.}
    \label{fig:V(r)}%
\end{figure}

\paragraph{Orbital Mechanics}
For a system bound by both cosmic strings and gravity, the radial effective potential is given by
\begin{equation}
    \label{eq:V(r)}
    U_J(r) = \mu r - \frac{R_1R_2}{r} + \frac{J^2}{2Rr^2},
\end{equation}
where $r$ is the separation between the binary's components. A sample is plotted in Fig.~(\ref{fig:V(r)}).
\noindent
The system only admits bound orbits, since ${U_J(r \to \infty) \propto r}$. Like in the previous subsections, any orbit is uniquely specified by its energy and angular momentum. For given $E$ and $J$, the orbit is bound by $r_p \leq r \leq r_a$, where the pericenter and apocenter radii are specified by equation $U_J(r)= E$. This gives a cubic equation in $r$ which for allowed energies has two positive roots. The negative root, as far as we are aware, has no physical significance.

\paragraph{Circular Inspiral}

For a circular orbit, $r_a=r_p$ and its evolution due to gravitational-wave emission can be computed in a straighforward manner. The energy and angular momentum decrease can be computed from Eqs.~(\ref{eq:eq1}) and (\ref{Icirc}). The angular frequency of rotation is given by 
\begin{equation}
    \omega=\sqrt{{\mu\over Ra}+{R_1+R_2\over a^3}}.
\end{equation}
Substituting this into the above equations, we get 
\begin{equation}
    \begin{aligned}
    & \frac{dE}{dt} = -\frac{32}{5} R^2 \left( \frac{R_1+R_2}{a^2} + \frac{\mu}{R} \right)^3 a \\
    & \frac{dJ}{dt} = -\frac{32}{5} R^2 \left( \frac{R_1+R_2}{a^2} + \frac{\mu}{R} \right)^{5/2} a^{3/2}
    \end{aligned}
\end{equation}

In the case of an orbit bound by cosmic strings and gravity, we have  
\begin{equation}
    E={3\over 2} \mu a-{R_1R_2\over 2a}.
\end{equation}
We obtain the decay rate for $a$ as 
\begin{equation}
    \label{eq: circular dr/dt}
    \frac{da}{dt} = \frac{da}{dE} \frac{dE}{dt} = - \lambda \frac{1}{a^3} \frac{(1+ \gamma a^2)^3}{1+3 \gamma a^2}
\end{equation}
for $\gamma \equiv \mu/(R_1 R_2)$ and $\lambda \equiv ({64}/{5})R_1 R_2 (R_1+R_2)$. We can exactly integrate this equation to obtain
\begin{equation}
    t = -\frac{1}{2 \lambda \gamma^2} \left( 3 \ln (1+\gamma a^2) + \frac{4+5\gamma a^2}{(1+\gamma a^2)^2} \right) + K,
    \label{K}
\end{equation}
where $K$ is a constant.
The decay of the circular orbit has two phases
\begin{enumerate}
    \item [i.] Early stage, string dominated. When ${\mu}/{F_g} = \gamma a^2 \gg 1$,  the polynomial factor is subdominant to the logarithmic factor; here $F_g=R_1 R_2/a^2$ is the force of gravitational attraction. We obtain $\ln(\gamma a^2) \simeq - 2 \lambda \gamma^2 t/3 + K'$, and $a(t) \simeq a_0 e^{-\lambda \gamma^2 t/3} = a_0 \exp[{-(64/15) \mu^2 t/R}]$, reproducing the behavior of pure string-driven inspirals, Eq.~(\ref{decay1}). 
    
    \item[ii.] Final stage, gravity-dominated. When $\mu/F_g = \gamma a^2 \ll 1$, we can expand the right-hand side of Eq.~(\ref{K}) in powers of $\gamma a^2$. We get, to the lowest order,
    \begin{equation}
        t-t_0 \simeq {1\over 4\lambda}(a_0^4-a^4),
    \end{equation}
    where the constants are chosen so that $a=a_0$ at $t=t_0$. 
    This is in agreement with the conventional gravitational circular inspiral formulae.
\end{enumerate}

\paragraph{General Orbital Decay}
A non-circular orbit bound by both gravity and a cosmic string is not closed by Bertrand's theorem, hence yielding analytic study very difficult. Instead, we analyze the decay of generic orbits numerically. 
First, we simulate the orbit by numerically integrating Newton's equations of motion
using the $4^{th}$-order Runge-Kutta method. We need to compute the motion over the full epicycle of the radial motion,  e.g.~from apocenter to apocenter. We tabulate a dense array of the traceless quadrupole moment tensor values along that orbit,  numerically differentiate and plug into Eq \ref{eq:eq1} to calculate the orbit-averaged rates of $\frac{dJ}{dt}, \frac{dE}{dt}$. We then employ a $2^{nd}$-order Runge-Kutta scheme where the orbit is evolved in the $J-E$ plane, and thus compute the evolution of $a,e$ with time.

\begin{figure}[H]%
    \centering
    \subfloat{{\includegraphics[width=8cm]{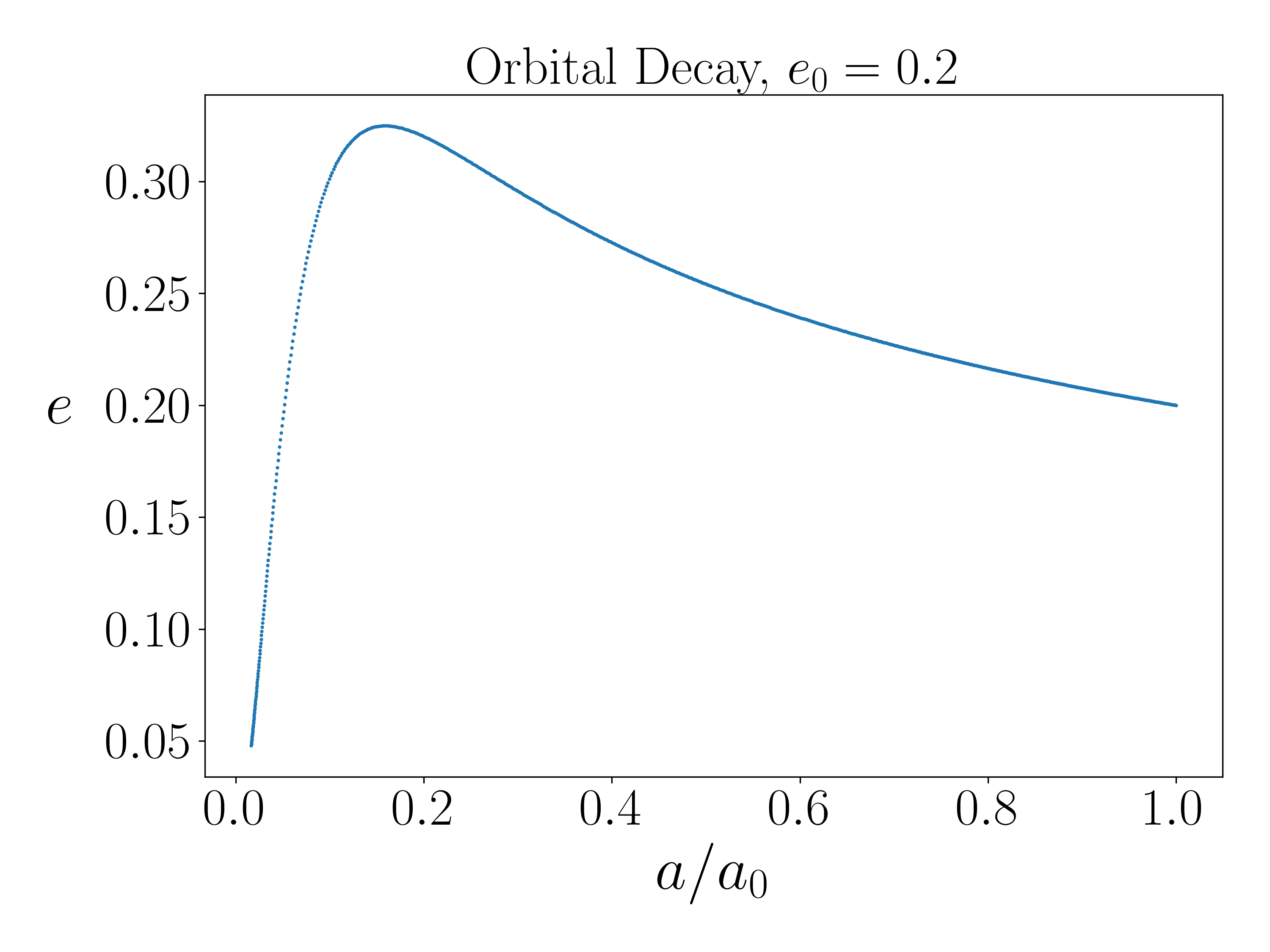} }}%
    \qquad
    \subfloat{{\includegraphics[width=8cm]{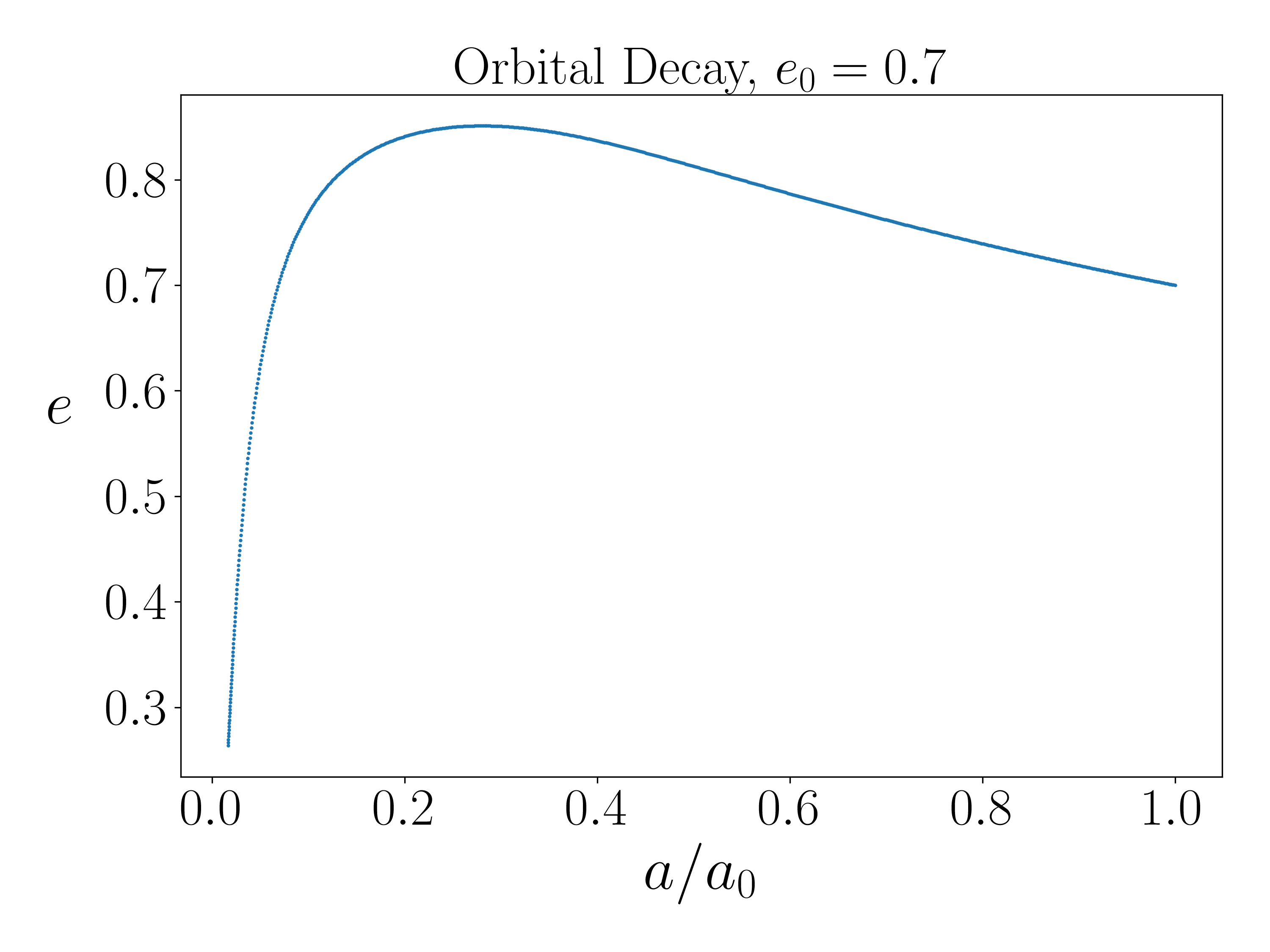} }}%%  
    \caption{Numerical calculation of the decay under both gravity and string tension, for $e_0=0.2, 0.7$, until $a(t)/a_0 < 1/60$, with $\mu R/(R_1+R_2) = 1/10$ and $\mu a_0^2/(R_1 R_2) = 90$.}
    \label{fig:ae}%
\end{figure}

\begin{figure}[H]%
    \centering
    \subfloat[$\mu R/(R_1+R_2) = 1/10$.]{{\includegraphics[width=8cm]{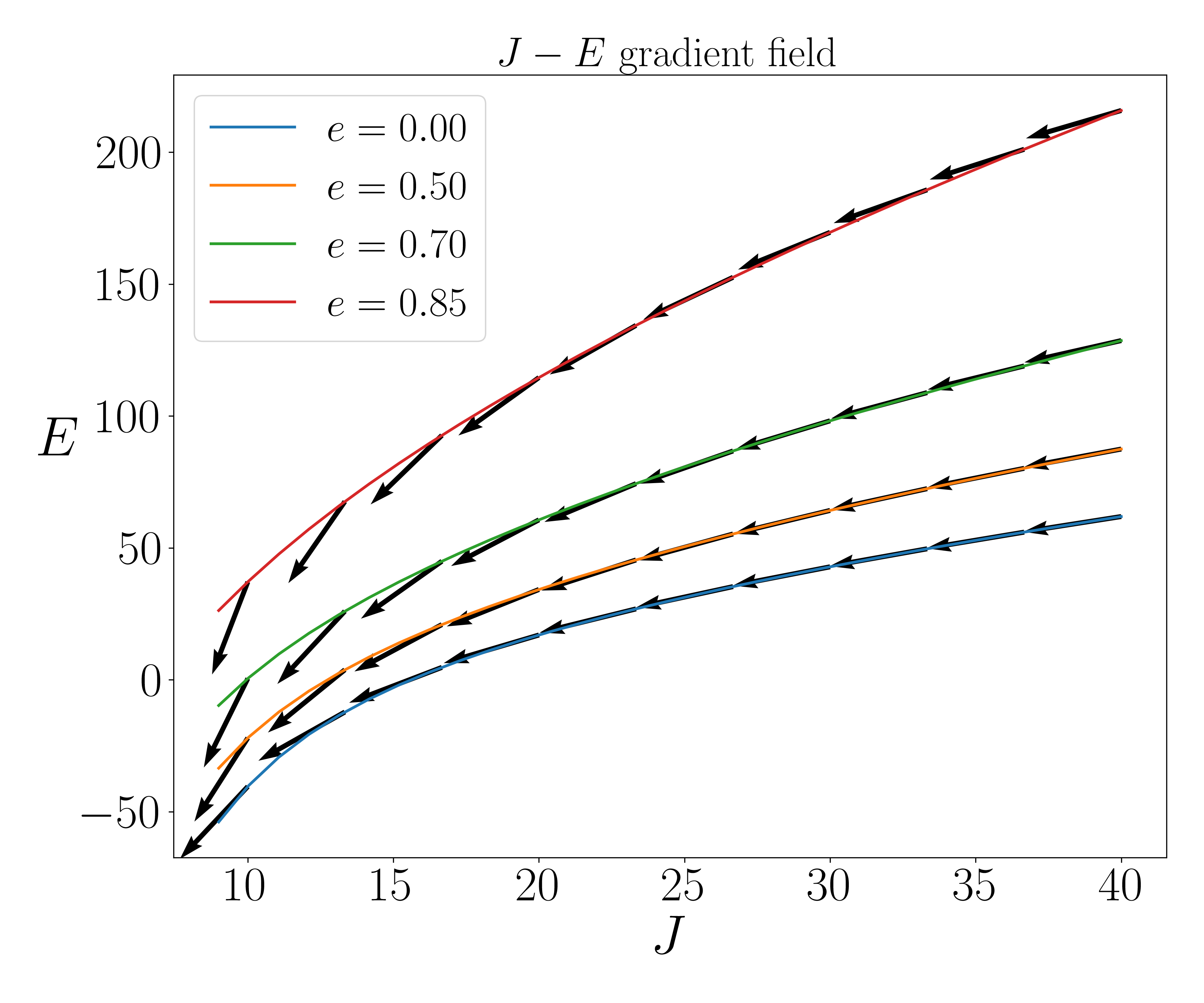} }}%
    \qquad
    \subfloat[$\mu R/(R_1+R_2) = 1$.]{{\includegraphics[width=8cm]{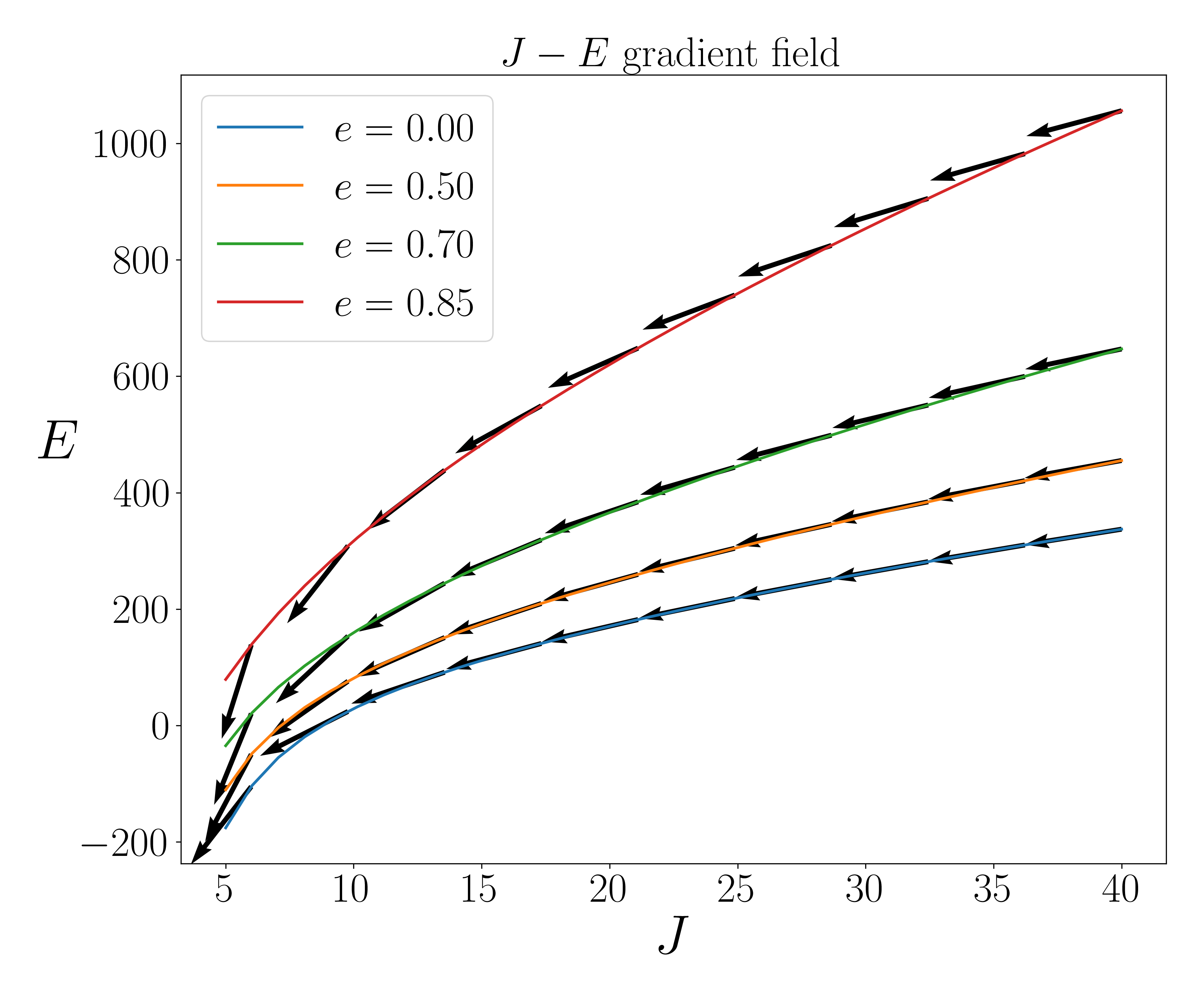} }}%%  
    \\
    \subfloat[$\mu R/(R_1+R_2) = 10$.]{{\includegraphics[width=8cm]{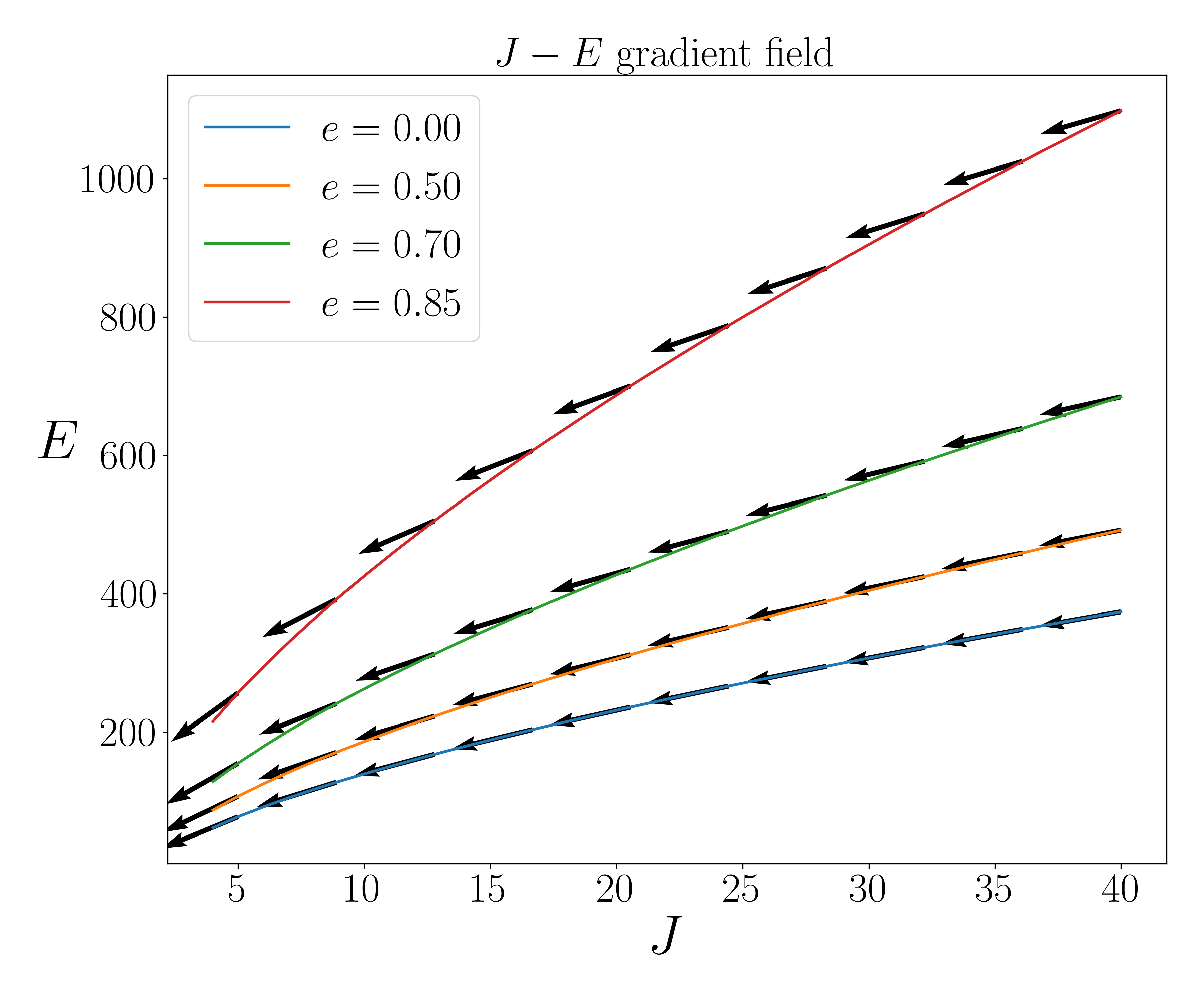} }}%
    \caption{Gradient field, $\left(\frac{dJ}{dt}, \frac{dE}{dt} \right)$, due to gravitational wave radiation. The lengths of vectors are non-uniformly scaled down, but the angles are preserved. In particular, the leftmost vectors are approximately scaled down by $\sim 10^8$, while the rightmost vectors are approximately scaled down by $\sim 10^4$.}
    \label{fig:El}%
\end{figure}

The flow vector fields in the $J-E$ plane due to gravitational wave radiation for a variety of cases are plotted in Fig.~(\ref{fig:El}). In particular, we note that for the early stages of the orbit in the string dominated regime (with high $J$ and high $a$), the eccentricity is generically driven to increase, more so for more eccentric orbits. Then, as the orbit decays and enters into the gravity dominated regime, it flows into less eccentric state, with the singular Keplerian rates as $a \to 0$ ensuring that the orbit decays in its final stages as a pure gravity-driven orbit, terminating in a circular state. Examples of this behavior are plotted in Fig.~(\ref{fig:ae}).

\section{Spins of black holes connected by a string} \label{sec:spins}
Vilenkin et al. \cite{2018JCAP...11..008V} noted that for a black-hole binary connected by a string, the gravitational-wave driven inspiral proceeds on a timescale shorter than a Hubble time for an interesting range of masses and string tension parameter. From Eq.~(\ref{decay1}), we see that the characteristic timescale for the exponential inspiral is given by
\begin{equation}
\tau={15 R\over 64\mu^2}\simeq 6\times 10^9 \left({M\over 10 M_\odot}\right)
\left({10^{-12}\over \mu}\right)^{-2}\hbox{yr}.
\end{equation}
Here $M$ is the reduced mass of the binary.
The actual time to merger is only a few times this value due to the super-exponential growth of eccentricity, regardless of the initial separation of the binary; see Eq.~(\ref{superexponential}). This calculation assumes that the binary is moving non-relativistically, i.e. that the mass of the string is less than the mass of the black holes, which translates into $a\lesssim R/\mu\simeq 5 (M/10 M_\odot)(10^{-12}/\mu)~\hbox{pc}$.  

However, the Peters' circularization of the binary should begin at a distance $a\sim R/\sqrt{\mu}=10^6(10^{-12}/\mu)^{1/2}~R$, large compared to the black-hole sizes. Therefore by the time the black holes merge, their orbits should be nearly circular. It is therefore interesting to ask whether there is another characteristic, apart from eccentricity, that will distinguish this type of merger. We argue below that the black holes that merge in this way are likely to have very low, but calculable spin values. The interest is purely academic at this point, since it is unlikely that such low spins would be meaningfully measured in any foreseeable experiment.

When an asymptotically straight string is attached to a spinning black hole of gravitational radius $r_0$, the latter experiences torque from the string. In the limit of slow rotation, the torque is given by 
\begin{equation}
{\bf Q}=-{4 \mu r_0^2}\left[{\bf \Omega}-({\bf \Omega}\cdot {\bf n}){\bf n}-{\bf n}\times\dot{\bf n}\right].
\label{Jevol}
\end{equation}
where $\bf \Omega$ is the angular velocity of the black hole's horizon, and $\bf n$ is the unit vector along the string at $r\gg r_0$ (closer to the black hole the string is dragged around it by the spin). This equation was first written by Xing et al.~\cite{2021PhRvD.103h3019X}, see their equation (4.19); it was based on the calculations of how stationary strings extract angular momentum from the black hole, which were first performed in \cite{1989PhLB..224..255F}. The validity of these expressions was tested by direct numerical experiments in \cite{2023PhRvD.107l3016D} \footnote{There is a long history of numerical studies of string motion in Kerr spacetime.  Early work ~\cite{Larsen_1994, Frolov_1999} considered the scattering of an axisymmetric, current-carrying string and showed that even such a simple system displays nontrivial chaotic behavior, while the later work~\cite{Snajdr_2002} considered the 3-dimensional scattering of a long string by the Kerr black hole.}.  We thus find that the angular velocity of the black hole evolves as follows:
\begin{equation}
 {\bf \dot{\Omega}}=-{\mu \over r_0}\left[{\bf \Omega}-({\bf \Omega}\cdot {\bf n}){\bf n}-{\bf n}\times\dot{\bf n}\right].  
 \label{stringspin}
\end{equation}

If the string is stationary relative to the black hole, we can neglect the $\dot{\bf n}$ term, and the black hole's spin only tends to align with the string, on the timescale $t_{\rm align}=r_0/\mu$, as explained in \cite{2021PhRvD.103h3019X}. However, in the case of a binary, the string cannot be considered stationary, since one can show that 
$P_{\rm orb}/t_{\rm align}\sim \sqrt{\mu a/R}\lesssim 1$ for a binary moving non-relativistically in the string-dominated phase. Hence, according to Eq.~(\ref{stringspin}), the black hole will also spin down to the orbital frequency, $\Omega \to 2\pi f$, on the same timescale $t_{\rm align}$; where $f = 1/P_{\rm orb} = \omega/2\pi$ is the orbital frequency. 

This alignment timescale is much shorter than the characteristic inspiral timescale in the string-dominated regime, ${t_{\rm align}/\tau\sim \mu \ll 1}$. Therefore in this inspiral stage, the black holes' spins become locked with the orbital frequency and are aligned with the string 
% orbital angular momentum; 
(perpendicular to the equatorial plane); the evolution equation during this stage of the inspiral simply becomes
\begin{equation}
\dot{\Omega}+{\mu\over r_0}\Omega={\mu\over r_0}2\pi f.
\end{equation}
For a circular inspiral this equation can be integrated exactly, but we find an estimate to be more instructive.

The black hole spins remain locked to the orbital motion until $\dot{f}/f\sim 1/t_{\rm align}= \mu/r_0$; they decouple when $\dot{f}/f \gtrsim 1/t_{\rm align}$. In Peters' regime,
\begin{equation}
{\dot{f}\over f}={96\pi^{8/3}\over 5} R_{\rm ch}^{5/3}f^{8/3},
\end{equation}
which means that the decoupling from the orbital motion will take place at the orbital frequency 
\begin{equation}
f_{\rm dec}=\Omega_{\rm dec}/(2\pi)\sim 0.1 \mu^{3/8} R_{\rm ch}^{-5/8} r_0^{-3/8}.
\label{fdec}
\end{equation}
Here $R_{\rm ch}$ is the chirp mass of the binary.
After decoupling, the black hole gets spun up according to 
\begin{equation}
\dot{\Omega}={\mu\over r_0} 2\pi f.
\end{equation}
The final spin of the black hole is given by
\begin{equation}
\Omega_{\rm merger}\sim\Omega_{\rm dec}+{\mu\over r_0} 2\pi N_{\rm orb},
\end{equation}
where $N_{\rm orb}$ is the number of orbits that the binary completes after decoupling and before the merger. The two terms on the right-hand side turn out to be of a similar order of magnitude. Therefore the final dimensionless spin parameter of the black hole is
\begin{equation}
s\sim \mu^{3/8} (r_0/R_{\rm ch})^{5/8},
\end{equation}
or, expressed in terms of the component masses, 
\begin{equation}
s_1\sim \mu^{3/8} (q+1)^{1/8} q^{1/4},
\end{equation}
where $q=R_1/R_2$ is the mass ratio of the two black holes. Clearly this is too small to be of any detectable significance, and will present as ``no spin" to all conceivable gravitational-wave measurements.

\section{Discussion} \label{sec:discussion}
The main result of this paper is that the eccentricity of a string-connected binary grows dramatically during the string-dominated phase of its inspiral. Initially exponential growth of eccentricity turns super-exponential, in a sense that when the quantity $1-e\ll 1$, it approaches zero much faster than exponential. 
The implications of this are as follows:

1. String-driven inspiral of monopole-antimonopole pairs will lead to head-on collisions that will likely destroy the string and annihilate the monopoles, when the orbital semimajor axis is still orders of magnitude greater than the string thickness. The direct collision will happen if the periastron distance is smaller than the string size. It is perhaps of interest to simulate such collisions directly, using the equations of motion for the fields that make the string, as opposed to using the Nambu-Goto action. 

2. String-driven mergers of light primordial black holes might be very eccentric; but the heavier stellar-mass black holes would circularize by Peter's mechanism. In all cases the black holes will be spun down to extremely low spins by the tension of strings attached to them. 

A major shortcoming of this work is that it does not treat the regime when the string is heavier than the binary components that are attached to it. In that case the binary motion becomes relativistic, and the approximations made in this work do not hold. This case will be considered in our future work.

 AS acknowledges support from the Black Hole Initiative at Harvard University, which is funded by grants from the John Templeton Foundation and the Gordon and Betty Moore Foundation. YL's work on this subject is supported by Simons Investigator Grant 827103.

\appendix

\section{Orbit averages} \label{sec:app averages}
The orbital period in Eq.~(\ref{period1}) can be computed by first casting it in a dimensionless form 
\begin{eqnarray}
    P(a,e)&=&2\int_{r_p}^{r_a}|v_r|^{-1}dr\\
        &=&2\sqrt{aR\over \mu}\int_{-1}^{1}{(ex+1)~dx\over \sqrt{(1-x^2)(3-e^2+2ex)}}\nonumber
\end{eqnarray}
Here we made a substitution $r=a(ex+1)$. The integral above was evaluated using Mathematica and the formulae in Eq.~(\ref{period1}) were checked by numerical integration of the above equation.

In both appendices, we restore $\langle \rangle$ in the notation to distinguish the orbit-averaged expression of an observable $\langle O \rangle$ from its local values during the orbit.

The simple-looking results in Eqs.~(\ref{rel1}) and (\ref{rel4}) can be obtained using the following considerations. From the Virial Theorem, for finite motion we have the following time-averages relationship:
\begin{equation}
\frac{1}{2}\langle \vec{F}\cdot\vec{r}\rangle+\langle E_k\rangle=0,
\end{equation}
where $\vec{F}$, $\vec{r}$, and $E_k$ are the generalized force, position, and kinetic energy of the system respectively. For the situation at hand this translates to 
\begin{equation}
    -\frac{1}{2}\mu \langle r\rangle+\langle E_k\rangle =0.
\end{equation} 
But the potential energy equals $\mu r$, so from the energy conservation we also have
\begin{equation}
\mu\langle r\rangle+\langle E_k\rangle=E.
\end{equation}
Combining the two, we get
\begin{equation}
\langle r\rangle={2\over 3}{E\over \mu}=a(1+e^2/3).
\end{equation}
This is Eq.~(\ref{rel1}). Furthermore, the radial equation of motion reads
\begin{equation}
    R\ddot{r}=-{\mu}+{J^2\over Rr^3}.
\end{equation}
Using the fact that $\langle \ddot{r}\rangle=0$,
we get
\begin{equation}
\left\langle{1\over r^3}\right\rangle={1\over a^3 (1-e^2)^2}.
\end{equation}
This is Eq.~(\ref{rel4})

The other two equations,~\eqref{rel2} and~\eqref{rel3}, were obtained by using the substitution $r=a(1+ex)$, performing the integration using Mathematica, and checking the results using numerical integration.

\section{Derivation of $f_E^\mu(e)$ and $f_J^\mu(e)$} \label{sec:app fE_fJ}
In this Appendix we sketch the derivation of the eccentricity dependence of the energy and angular momentum losses, as expressed through the form-factors $f_E^{\mu}(e)$ and $f_J^\mu(e)$. Our starting point are the standard expressions for the quadrupole-driven radiation reaction:
\begin{equation}
    \label{losses}
    \begin{aligned}
    & \left\langle \frac{dE}{dt} \right\rangle = -\frac{1}{5} \left\langle \frac{d^{3} I_{ij}}{dt^3}\frac{d^{3} I_{ij}}{dt^3} \right\rangle, \\
    & \left\langle \frac{dJ_{i}}{dt} \right\rangle = -\frac{2}{5} \epsilon_{ijk} \left\langle \frac{d^2 I_{jm}}{dt^2} \frac{d^3 I_{km}}{dt^3} \right\rangle.
    \end{aligned}
\end{equation}
The quadrupole moment of the binary is given by 
\begin{equation}
\frac{1}{R} I_{ij}= x_i x_j-{1\over 3}r^2 \delta_{ij}.
\label{standardQ}
\end{equation}
Here, as in the main text, $R$ is the reduced mass of the binary, and $\vec{r}=(x_1, x_2, x_3)$ is the vectorial separation of the binary's members. The equations of motion of the binary are
\begin{eqnarray}
    {dx_i\over dt}&=&v_i\nonumber\\
    {dv_i\over dt}&=&-\frac{\mu}{R}{x_i\over r}.
    \label{motion}
\end{eqnarray}
% where $g=\mu/R$. 
The angular momentum and energy of the binary are 
\begin{eqnarray}
J_i&=&R\epsilon_{ijk}x_i v_j\nonumber\\
E&=& {1\over 2}R v^2+\mu r.
\label{amenergy}
\end{eqnarray}
Also it is convenient to define radial and tangential velocities,
\begin{eqnarray}
    v_r&=&\vec{v}\cdot \vec{r}/r= \dot{r},\\
    v_t&=&\sqrt{v^2-v_r^2}= {J\over r R}.
    \label{vt}
\end{eqnarray}

Repeatedly differentiating Eq.~(\ref{standardQ}) with respect to time, and using the equations of motion, we obtain the following:
\begin{eqnarray}
{1\over R}{dI_{ij}\over dt}&=& v_ix_j+x_iv_j-{2\over 3} r v_r \delta_{ij},\\
{1\over R}{d^2I_{ij}\over dt^2}&=& 2  \left( {\mu \over R} {r\over 3} \delta_{ij}-{\mu \over R}{x_i x_j\over r}+v_i v_j-{v_r^2\over 3}\delta_{ij}\right),\nonumber\\
{1\over R}{d^3I_{ij}\over dt^3}&=& 2{\mu \over R}\left[{v_r x_i x_j\over r^2}+v_r\delta_{ij}-{2\over r}(v_i x_j+ v_j x_i)\right]. \nonumber
\end{eqnarray}
Using these expressions, we obtain
\begin{equation}
    {d^3 I_{ij}\over dt^3}{d^3 I_{ij}\over dt^3}=\mu^2 (24 v_r^2+ 32 v_t^2),
    \label{firstsum}
\end{equation}
and 
\begin{equation}
    {d^2 I_{jm}\over dt^2}{d^3 I_{km}\over dt^3}=4\mu R\left[2{\mu \over R}+{v_r^2\over r}+2{v_t^2\over r}\right] x_j v_k+ S_{jk}.
    \label{secondsum}
\end{equation}
Here $S_{jk}$ is symmetric with respect to indices $j,k$, and therefore it gives zero when contracted with $\epsilon_{ijk}$. We thus have
\begin{equation}
    \label{losses1}
    \begin{aligned}
    &  \frac{dE}{dt}  = -\frac{\mu^2}{5} (24 v_r^2+32 v_t^2), \\
    & \frac{d\vec{J}}{dt}  = -\frac{8\mu}{5} \left[2{\mu \over R}+{v_r^2\over r}+2{v_t^2\over r}\right]\vec{J}.
    \end{aligned}
\end{equation}
To obtain Eqs~(\ref{eccentricityfactors}), 
(\ref{eq:fE}), and (\ref{eq:fJ}), we need to orbit-average the above equations. To achieve this, we first use 
\begin{equation}
    {v_r^2}={2E\over R }- \frac{2 \mu r}{R}- {J^2\over R^2 r^2}
\end{equation}
and Eq.~(\ref{vt}) to express the right-hand side of Eq.~(\ref{losses1}) in terms
of $r$, $1/r$, $1/r^2$, and $1/r^3$. Averages of the latter are given in Eqs.~(\ref{rel1}) --- (\ref{rel4}). Using those together with the expressions in Eq.~(\ref{eq:k=0, El}), after some amount of algebra we arrive at Eqs.~(\ref{eccentricityfactors}), (\ref{eq:fE}), and (\ref{eq:fJ}).

\bibliography{main}

\end{document}